\pdfoutput=1
\documentclass[a4paper,12pt]{article}
\usepackage[english]{babel}
\usepackage{amsmath}
\usepackage{graphicx}
\usepackage{graphics}
\usepackage[font=small]{caption}
\usepackage{multirow}
\usepackage[tiny]{titlesec}
\begin{document}
\title{A thermodynamic analysis of the spider silk and the importance of
complexity}
\small{{\author{s. ripandelli$^{*\alpha}$, d. pugliese$^{\alpha}$ and u. lucia$^{\beta}$}}}
\maketitle
\small{
$^{\alpha}$ Dipartimento di Scienza Applicata e Tecnologia, Politecnico
di Torino, Corso Duca degli Abruzzi 24, 10129, Torino (Italy)\\
$^{\beta}$ Dipartimento di Energia, Politecnico di Torino, Corso
Duca degli Abruzzi 24, 10129, Torino (Italy)\\\\
{$^{*}$corresponding} author: simone.ripandelli@gmail.com}
\section*{Abstract }
The spider silk is one of the most interesting bio-materials investigated
in the last years. One of the main reasons that brought scientists
to study this organized system is its high level of resistance if compared to other artificial materials
characterized by higher density. Subsequently, researchers discovered
that the spider silk is a complex system formed by different kinds
of proteins, organized (or disorganized) to guarantee the required
resistance, which is function of the final application and of the
environmental conditions. Some spider species are able to make different
silks, up to twelve, having a composition that seems to be function
of the final use (i.e. dragline web, capture web, etc). The aim of
this paper is to analyze the properties of the spider silk by means
of a thermodynamic approach, taking advantage of the well-known theories
applied to polymers, and to try to underline and develop some intriguing
considerations. Moreover, this study can be taken as an example to
introduce and discuss the importance of the concept of optionality
and of the anti-fragile systems proposed by N. N. Thaleb in his book
\textquotedblleft Antifragile: Things that gain from disorder\textquotedblright . 
\section{General Introduction }
The present work can be divided in three sections. In the first one,
a general description of the silk using the olog approach$^{[1]}$
is proposed. This part is useful to schematize the possible behaviour
of the spider silk. In the second section, the hierarchical nature
of the spider silk is investigated through a thermodynamic approach.
The right side of Figure 2.1 shows two examples of the spider web,
while in the left side of the picture a scheme of the hierarchical
spider silk structure is depicted. It was the starting point to build
up the description proposed. 
\begin{figure}[htb]
\centering
\includegraphics[scale=0.5]{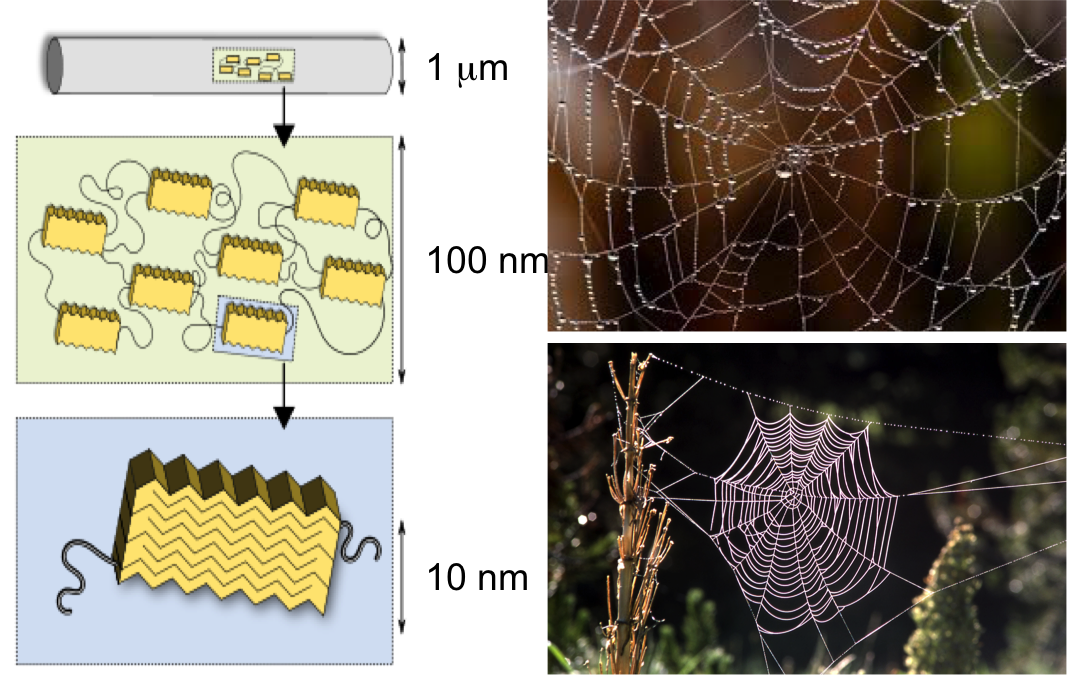}
\caption{On the left, a spider web; on the right, a schematic picture of the
spider silk structure $^{[3]}$.}
\end{figure}
Finally, in the third section, it has been discussed how the interpretation
proposed can be taken as an example to explain the importance of the
antifragile theory of N. N. Thaleb$^{[2]}$, where the antifragility
should not be interpreted only as a toughness anymore. From another
point of view, the interpretation proposed suggests that the nature
follows the ideas of complexity and optionality: maybe the most important
keys to survive. The olog approach was considered necessary to create
an initial and general scheme of the structure of the spider silk
(see Figure 2), putting in evidence important aspects necessary to
build up the proposed interpretation. 
\begin{figure}[htb]
\centering
\includegraphics[scale=0.5]{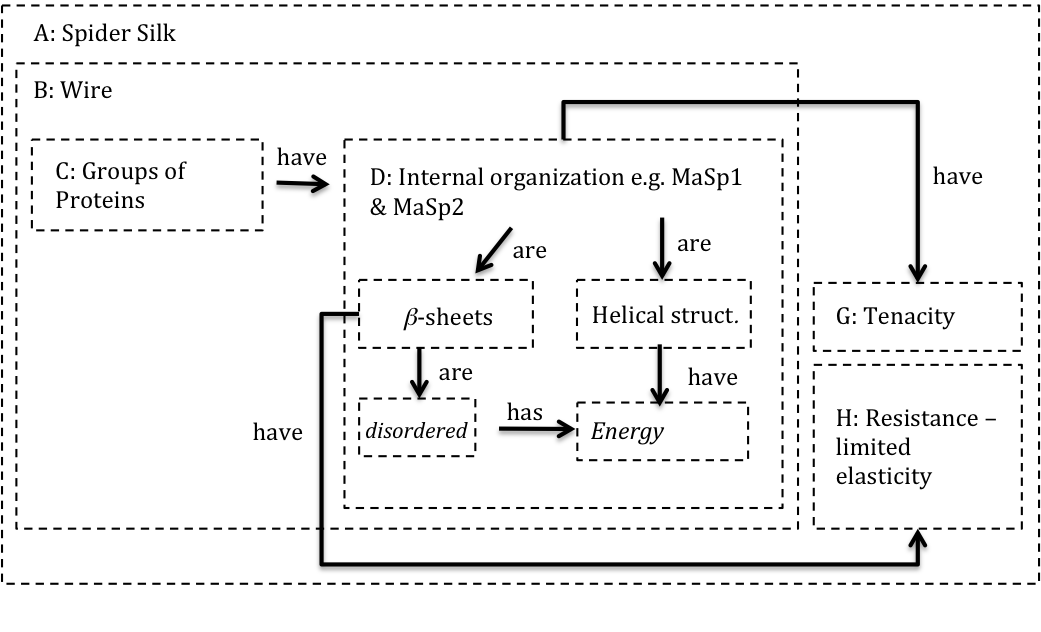}
\caption{General possible olog scheme of the spider silk behaviour.}
\end{figure}
In order to approach the problem from a thermodynamic point of view,
the starting point taken into the account was the equation of free
energy proposed by Helmholtz (2.1):
\begin{equation}
dA=-TdS+\mu dN+\sigma d\epsilon
\end{equation}
The equation (2.1) collects all the terms that describe \textquotedblleft where\textquotedblright{}
the energy is stored in a system. The free energy (F) represents the
work that the system is able to do on the environment or, from another
point of view, the energy that the system can provide when it is stressed
by an external force. In order to use the equation of the free energy,
it was necessary to identify all possible actors that play a role
in the \textquotedblleft energy storage\textquotedblright . Referring
to the general scheme of the internal structure of the spider silk
(see Figure 1) and starting from the internal structure deeply studied
in literature at the level of protein composition $^{[4,5]}$, different
responses can occur at the microscopic level when a spider web is
stressed. 
Upon the occurrence of an external stress, the following phenomena
were considered at the microscopic level:
\begin{itemize}
\item the crystalline part of the silk, formed by $\beta$-sheets based
proteins, becomes oriented on the force direction, thus reaching an
internal order (i.e. the rectangles represented in Figures 1 and 3).
\item the amorphous phase, that links the $\beta$-sheets structures, is
stretched and the hydrogen bonds (H-H) between these glycine chains
are broken. At the same time, the distance between the $\beta$-sheets
increases and the system starts to work as a spring.
\item the same behaviour of the $\beta$-sheets occurs at the amorphous
bonds level, i.e. the long chains of glycine are stretched.
\end{itemize}
After the third step, and before to reach the rupture, the system
may offer a final resistance (proportional to the Young\textquoteright s
modulus E$_silk$ of the entire system). Anyway, in this study this
behaviour was evaluated negligible and therefore was not considered.
After these considerations, it is possible to re-write the equation
(2.1) as follows: 
\begin{equation}
A=-T(S_{\vartheta}+S_{M}+S_{m})+\mu\Theta+\sigma\epsilon
\end{equation}
The last term on the right side was neglected, and the three entropic
terms were differently approached.
The first entropic term, S$_{teta}$, refers to the $\beta$-sheets
orientation and was considered as a first response of the material.
Having the possibility to re-organize its internal structure, the
spider silks re-arranges itself in order to exhibit the maximum level
of resistance. It passes from a disordered phase to an ordered one.
From a theoretical point of view, part of the internal energy stored
is thus employed to change its configuration or, from another point
of view, it is initially stored into the disordered configuration
(i.e. the spiders exploit the disorder). Looking for a representation
of this behaviour, the literature provided an exhaustive scheme of
the spider silk response (see Figure 3)$^{[6,7]}$. In particular,
the phenomenon of $\beta$-sheets orientation
is well depicted passing from panels g to h, before and after the
stretching.
\begin{figure}[htb]
\centering
\includegraphics[scale=0.75]{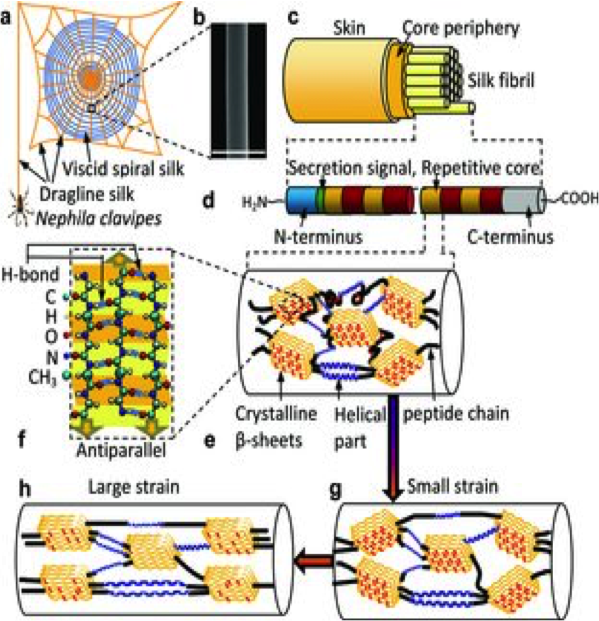}
\caption{Scheme of the spider silk response.}
\end{figure}
\section{Thermodynamic Formulation}
The entropic term due to the initial disorder can be written using
the Boltzmann\textquoteright s relation (3.1):
\begin{equation}
S_{\vartheta}=-k_{B}lnW
\end{equation}
where $k_{B}$ is the Boltzmann\textquoteright s constant and $W$
is the configuration assumed by the system. The possible mobility
of the $\beta$-sheets in a plane was considered to describe the configuration
of the system. If a $\beta$-sheet is not oriented in the stress direction,
it rotates up to reach the longitudinal direction. In other words,
the energy stored as disposition with respect to the longitudinal
direction is utilized at this level. In light of this aspect, the
configuration entropy was used considering the possibility to have
many small systems able to rotate and having initial directions in
the range {[}-$\pi$/2, $\pi$/2{]}; where 0 degrees is the condition
when the $\beta$-sheets are aligned. Therefore, a $\beta$-sheet
can assume two positions per rotation having the same energy quantity.
Re-writing the equation (3.1), it is possible to assess that:
\begin{equation}
W(\vartheta)=\sum_{i}p_{i}lnp_{i}=\sum_{i}n_{i}e^{-\frac{E_{i}}{Tk_{B}}}ln\left(n_{i}e^{-\frac{E_{i}}{Tk_{B}}}\right)
\end{equation}
where $n_{i}$ is the number of the states and $E_{i}$ is the energy
stored. Taking into the account that the energy is a force per displacement,
it is possible to consider a total rotation of the $\beta$-sheets
when an external stress is applied. In particular: 
\begin{equation}
E_{i}=F_{i}\vartheta d=Al_{\%}F_{ext}cos(\pi/2-\vartheta)\vartheta d
\end{equation}
where ${Al_\%}$ is the percentage of the entire volume composed by Alanine, which is the main component of $\beta$-sheet structures, $F_{ext}$ is the external force exerted and d is the entire length of the piece of web analyzed. By substituting the equation (3.3) in the equation (3.2), it is possible to obtain the equations (3.4): 
\begin{equation}
W(\vartheta)=\sum_{i}p_{i}lnp_{i}=\sum_{i}n_{i}e^{-\frac{Al_{\%}F_{ext}cos(\pi/2-\vartheta)\vartheta d}{k_{B}T}}ln(n_{i}e^{-\frac{Al_{\%}F_{ext}cos(\pi/2-\vartheta)\vartheta d}{Tk_{B}}})
\end{equation}
or (3.5),
\begin{equation}
S_{\vartheta}=-k_{B}lnW(\vartheta)
\end{equation}
The other two entropic terms were obtained starting from a thermodynamic
model $^{[8-11]}$ commonly used to describe the behaviour of polymers.
The idea was to employ the theory separately, describing the silk
at two levels and building up two different entropic terms. Indeed,
observing the model proposed for the spider silk, it can be intuitively
noticed that before the breakage of all the H-H bonds, and immediately
after the alignment of the $\beta$-sheets, the silk starts to be
stretched and its internal rigid components are subjected to a sort
of deformation that augments the distance between them. A general
scheme used to describe this theoretical behaviour is depicted in
Fig. 3.1.
\begin{figure}[htb]
\centering
\includegraphics[scale=0.75]{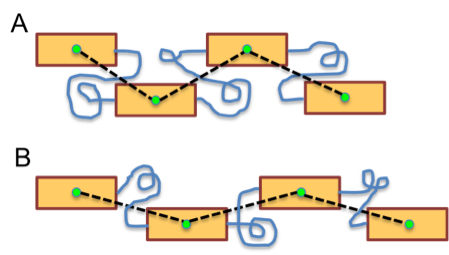}
\caption{Macrostructure where the $\beta$-sheets are linked one to each other
(A). The structure, simplified with nodes and bonds, undergoes a relaxation
when an external stress is applied (B)}
\end{figure}
At this level no breakage of bonds occurs, and the entropy varies
because of the changes in the organization of the structure. The entropic
term describes the passage from an higher level of disorder to a lower
one, with the occurrence of a transformation. The second and third
entropic terms were built up starting from the same theory explained
above but applied to a lower scale, where the H-H bonds in the amorphous
phase are broken and another step of relaxation occurs. Referring
to the Fig. 3.1, it is possible to put in evidence different aspects.
The scheme of the structure (i.e. nodes and bonds) are composed by
bonds with the same length, $\textit{b}$, that link not contiguous species.
In particular, the end-to-end distance is: 
\begin{equation}
R_{E}=\sum_{i=1}^{N}\bar{r_{i}}
\end{equation}
\begin{equation}
\left\langle R_{E}^{2}\right\rangle =Nb^{2}
\end{equation}
This result follows from the ones obtained by Khun on the rubber $^{[8]}$.
Passing to a lower level, the third entropic term was analyzed with
a similar approach, the freely joined chains theory (FJC)$^{[12]}$,
which was considered much more significant to describe the system.
Carrying out the calculus, it is possible to extrapolate two potentials
related to the \textquotedblleft macro\textquotedblright{} and \textquotedblleft micro\textquotedblright{}
applications of the theory. 
\begin{equation}
TS_{M}=k_{B}T\hphantom{}lnp(M)=k_{B}T\hphantom{}H(M)=\frac{3}{2}\frac{k_{B}T}{b}\sqrt{N\hphantom{}Al_{\%}}
\end{equation}
\begin{equation}
TS_{M}=k_{B}T\hphantom{}lnp(m)=k_{B}T\hphantom{}h(m)=\frac{3}{2}\frac{k_{B}T}{b^{\frac{3}{2}}}\sqrt{N\hphantom{}}
\end{equation}
In the first relation, an important role is played by the Al$_{\%}$
term, which is the percentage of the volume able to perform this macro-behaviour.
Indeed, the entropic term $\textit{SM}$ was built with the aim to describe
the macroscopic behaviour performed by the $\beta$-sheets components.
Once all the entropic terms reported in equation (2.2) have been thoroughly
discussed, the attention can be now focused on the term $\mu\Phi$,
which describes the energy stored in the H-H bonds between the proteins.
The number of these bonds should not be fixed. Indeed, as the silk
reology is strictly related to the RH level of the environment and
as by nature the silk has an hydrophobic behaviour, it is possible
that exposing a smaller surface to a higher humidity level the contraction
of the structure could bring to the formation of a higher number of
H-H bonds. Referring to the literature and on the experiments performed$^{[13,14]}$,
the resistance of the spider silk is strictly dependent on the relative
humidity of the air. Intuitively, in conditions of high relative humidity
(RH) the intrinsic hydrophobic nature of the silk leads to a contraction
of the web and thus to a higher storage of energy. More specifically,
in presence of water, even if from one side the silk is able to increase,
at time 0, the quantity of energy stored, on the other hand there
is a higher possibility for the proteins to restore H-H bonds with
water molecules. This theory is supported by the fact that when the
humidity is higher than a certain limit, the Young\textquoteright s
module decreases and the resilience response of the silk increases.
Referring to this interpretation, the energy stored and in particular
the number of bonds are strictly related to the RH level:
\begin{equation}
\mu\propto RH\;or\;\mu(HR)
\end{equation}
It is impossible to know exactly the location and the number of the
bonds present in each structure. For this reason, the approach chosen
is also for this case probabilistic. It is possible to define the
number of bonds $\mu$ as follows, in function of the RH value:
\begin{equation}
\mu(HR)=\frac{1}{\sqrt{2\pi\sigma^{2}(RH)}}e^{-\frac{\mu^{2}}{2\sigma^{2}(RH)}}
\end{equation}
The number of bonds at different RH conditions should be determined
through experimental tests. The choice to employ this kind of distribution
came out from the experimental evidences reported by Vehoff and co-workers$^{[5]}$.
Once to have explicated all the terms, the equation (2.2) can be re-written
as follows:
\begin{equation}
\begin{split}
A& =-k_{B}T\hphantom{}ln\left\{ n\left[\sum_{i}n_{i}e^{-\frac{Al_{\%}F_{ext}(\pi/2-\vartheta_{i})\vartheta_{i}d}{Tk_{B}}}ln\left(n_{i}e^{-\frac{Al_{\%}F_{ext}(\pi/2-\vartheta_{i})\vartheta_{i}d}{Tk_{B}}}\right)\right]\right\} +\\
 & \quad \frac{3}{2}\frac{k_{B}T}{b}\sqrt{Al_{\%}N}+\frac{3}{2}\frac{k_{B}T}{b^{\frac{3}{2}}}\sqrt{N}+\varTheta\left(\frac{1}{\sqrt{2\pi\sigma^{2}(RH)}}e^{-\frac{\mu^{2}}{2\sigma^{2}(RH)}}\right)
\end{split}
\end{equation}
From this equation, it is finally possible to extrapolate the value
of the force:
\begin{equation}
F=\frac{dA}{dx}
\end{equation}
Experimental tests and comparison with other results taken from the
literature will be conducted in order to calculate the value of equation
3.11.
\section{Conclusion}
This study deals with the simple description of a natural material
such as the spider silk. The main goal of the present work is to start
from the natural complexity of the spider silk to put in evidence
a general formulation of its mechanical behaviour, useful to design
new materials with the same operating principle$^{[15]}$. From another
point of view, this study wants to pose some considerations on how
the complexity in collaborative systems is one of the prerogative
to survive. This aspect is at the base of the anti-fragile theory
and of its peculiar optionality characteristic. The optionality of
the spider silk could be considered as an interesting example and
supports the theory of N. N. Thaleb. One of the most fascinating aspects
of the present study on the spider silk lies within its particular
nature. Indeed, it seems that spiders had well understood how to exploit
natural principles such as entropy and strength of collaborating systems$^{[16]}$.
When a spider makes a web, it has not the possibility to forecast
all the different conditions in which the web will operate. The lack
of this possibility makes the spider to develop a strategy for building
up a versatile material, able to store energy at different scales,
that interact each other, and to resist at different conditions. A
social example can be extrapolated from this consideration: a social
multicultural community such as the one in a small company in the
R{\&}D department able to face unpredictable events. For this reason,
the spider silk can be considered as an anti-fragile material and
a good example of social anti-fragile communities or behaviours. It
is not so tough to resist to a specific external stress, but it is
tough enough to learn from the environment, rearranging itself in
function of the humidity and becoming able to resist at different
external stresses. This last capacity comes from the possibility of
the silk to adapt its structure: from a certain point of view, the
silk learns and reorganizes its components once the environmental
conditions taught to it in which conditions it is going to operate.
The social communities should operate following an equivalent approach.
Even if we are comparing a thinking being with an inanimate material,
one more time the nature is teaching us how the forecast is much less
important if we learn to be prepared to face all possible conditions.
On the other hand, this is a prerogative of the wise people.

\end{document}